\documentclass[a4paper,9pt]{article}

\usepackage{INTERSPEECH2020}
\usepackage{amsmath,amssymb,bm,pgfgantt,enumitem}
\usepackage[]{hyperref}
\hypersetup{
    linkcolor=black
}
\usepackage[colorinlistoftodos]{todonotes}
\usepackage{caption,subcaption,multirow,cleveref}

\usepackage[ruled,vlined,linesnumbered]{algorithm2e}
\usepackage{float}
\usepackage{cleveref}
\usepackage{tikz,pgfplots}
\title{Approximate Fixed-Points in Recurrent Neural Networks}
\name{Zhengxiong Wang and Anton Ragni}
\address{
Department of Computer Science, University of Sheffield, 211 Portobello, Sheffield S1 4DP, UK}
\email{donatello.wang@outlook.com, a.ragni@sheffield.ac.uk}

\begin{document}

\maketitle
\begin{abstract}
Recurrent neural networks are widely used in speech and language processing. Due to dependency on the past, standard algorithms for training these models, such as back-propagation through time (BPTT), cannot be efficiently parallelised. Furthermore, applying these models to more complex structures than sequences requires inference time approximations, which introduce inconsistency between inference and training. This paper shows that recurrent neural networks can be reformulated as fixed-points of non-linear equation systems. These fixed-points can be computed using an iterative algorithm exactly and in as many iterations as the length of any given sequence. Each iteration of this algorithm adds one additional Markovian-like order of dependencies such that upon termination all dependencies modelled by the recurrent neural networks have been incorporated. Although exact fixed-points inherit the same parallelization and inconsistency issues, this paper shows that approximate fixed-points can be computed in parallel and used consistently in training and inference including tasks such as lattice rescoring. Experimental validation is performed in two tasks, Penn Tree Bank and WikiText-2, and shows that approximate fixed-points yield competitive prediction performance to recurrent neural networks trained using the BPTT algorithm. 

\end{abstract}

\section{Introduction}
Recurrent neural networks (RNN) are a popular choice for solving a variety of natural language and speech processing tasks including machine translation \cite{sutskever2014}, language modelling \cite{mikolov2010} and acoustic modelling \cite{graves2012, chan2016, peddinti2017low}.
These powerful highly non-linear sequence models enable to model long-range dependencies impossible with many other model classes. 
Although RNNs have achieved excellent performance across many applications and tasks their practical use comes with a range of well-known issues \cite{pascanu2013}.
As such, a variant of RNNs \cite{Hochreiter1997, cho2014learning}
or a completely alternative sequence model \cite{vaswani2017} that can overcome those issues are of both practical and theoretical interest. 

Unlike simpler $n$-gram and feed-forward neural network models used in language modelling \cite{schwenk2012continuous}, RNNs make predictions based on the complete rather than truncated history. The complete history representation used by RNNs is known to be unstable, which may cause optimisation issues and loss of information about long-term dependencies \cite{bengio1994learning}.
Although alternative history representations have been proposed to address these issues \cite{Hochreiter1997, cho2014learning},
the robust and accurate modelling of long-term dependencies in these models remains a challenge. Long-term dependency modelling also causes issues \cite{chen2014efficient} 
for efficient parallelization of training these models using algorithms such as back-propagation through time (BPTT) \cite{werbos1990backpropagation}, which leads to increased development costs and time. Although computationally efficient training methodologies have been proposed \cite{williams1990efficient}
to replace BPTT, they do not offer easier parallelization. Furthermore, many important applications of RNNs involve rescoring graph-like lattice structures, which, unlike linear chains or prefix trees, require inference-time approximations \cite{liu2014efficient, xu2018pruned}.
%

This paper shows that the form of RNNs used in language modelling yields history representations that can be reformulated as fixed-points of a non-linear equation system. These fixed-points can be computed either exactly or approximately using an iterative algorithm with guaranteed convergence in as many steps as the lengths of underlying sequences. Each iteration in this algorithm enables to account for one additional order of dependencies until at termination the dependencies of all orders have been accounted for. Although approximate fixed points have fewer dependencies than provided by standard RNN history representations, they have a number of interesting properties. First, according to Banach theorem \cite{zhou2018graph}, 
approximate fixed points converge to exact fixed points exponentially fast which supports terminating the iterative algorithm after few iterations. Second, unlike standard RNN history representations, approximate fixed points can be parallelised. Third, given the ability to control the scope of dependencies (the number of iterations), the approximate fixed-points can be applied for lattice rescoring without making any inference time approximations.

The rest of this paper is organised as follows. Section \ref{section:Recurrent Neural Networks} describes the form of RNNs examined in this paper. Section \ref{section:Fixed-Point Representation} shows how RNN history representations can be reformulated as fixed-points and presents an algorithm for computing them by iteratively refining approximate fixed-points. Experimental results comparing the RNN history representations, exact and approximate fixed points on two language modelling tasks are presented in Section \ref{section:Experiments}. Finally, conclusions drawn from this work are given in Section \ref{section:Conclusion and Future Work}.

\section{Recurrent Neural Networks}
\label{section:Recurrent Neural Networks}
There are many variants of recurrent neural networks (RNNs) \cite{elman1990finding, Hochreiter1997, cho2014learning, jordan1997serial, BiRNN1997}.
This paper examines the class of RNNs commonly referred to as Elman networks \cite{elman1990finding}. 
An Elman network consists of an input layer, a hidden or history layer and an output layer. The hidden layer computes a history representation at time $t$ using the following recursive process\footnote{For simplicity all biases are omitted in the exposition.}
\begin{equation}
    {\bm h}_{t} = {\bm\phi}({\bm W}{\bm h}_{t-1} + {\bm V}{\bm x}_{t-1})\label{eq:elman}
\end{equation}
where ${\bm x}_{t-1}$ is an one-hot encoding or an embedding \cite{pennington2014glove, mikolov2013efficient}
of a previous input word $w_{t-1}$, ${\bm h}_{t-1}$ is a previous history representation, ${\bm\phi}$ is an activation function, such as {\em{tanh}} and {\em{sigmoid}}, ${\bm V}$ and ${\bm W}$ are weight matrices associated with the input and hidden layer respectively. The history representation is used by the output layer to compute probability of predicting next input word, $w_{t}$, using {\em softmax} activation function
\begin{equation}
    P(w_{t}=i|{\bm h}_{t};{\bm\theta}) = \frac{\exp({\bm u}_{i}^{\top}{\bm h}_{t})}{\sum_{j=1}^{|V|} \exp({\bm u}_{j}^{\top}{\bm h}_{t})}
    \label{eq.pro of word i}
\end{equation}
where ${\bm u}_{i}$ is the $i$-th row of output layer weight matrix ${\bm U}$, $|V|$ is the size of vocabulary and ${\bm\theta}$ is the set of RNN parameters. The `quality' of RNN predictions is assessed using perplexity 
\begin{equation}
    \text{PPL}({\mathcal D};{\bm\theta}) = \exp\left(-\frac{1}{|{\mathcal D}|}\sum_{r=1}^{R} \log P({\bm w}_{1:T_{r}}^{(r)};{\bm\theta})\right)
\end{equation}
where ${\mathcal D}$ is a held-out set of $R$ test word sequences and $|{\mathcal D}|$ denotes the number of words. The dependency on the complete past input sequence implied by~\cref{eq:elman} makes it challenging to apply RNNs to structures other sequences. For instance, in order to apply RNN language models for rescoring lattices in speech recognition, a range of inference time approximations is commonly used \cite{liu2014efficient, xu2018pruned}.


The RNN parameters ${\bm\theta}$ can be estimated by minimising cross-entropy loss function
\begin{eqnarray}
    {\mathcal L}({\mathcal D};{\bm\theta}) &=& -\frac{1}{R}\sum^R_{r=1}\log(P({\bm w}^{(r)}_{1:T_r};{\bm\theta}))\\
    &=& -\frac{1}{R}\sum^R_{r=1} \sum_{t=1}^{T_r} \log(P(w_{t}^{(r)}|{\bm h}_{t}^{(r)};{\bm\theta}))
    \label{eq.CE classification}
\end{eqnarray}
 where $\mathcal D$ is a training set of $R$ word sequences. For optimisation it is common to use stochastic gradient-based schemes \cite{bottou2010large}.
 In such cases ${\mathcal D}$ is a sample (mini-batch) drawn from the training set. 
The gradient of the cross-entropy loss function with respect to the RNN parameters is
\begin{equation}
    \frac{\partial {\mathcal L}({\mathcal D};{\bm\theta})}{\partial {\bm\theta}} \propto
    \sum_{r=1}^{R}
    \sum_{t=1}^{T_r}
    \sum_{k=1}^{t} 
    \frac{\partial\log(P(w_{t}^{(r)}|{\bm h}_{t}^{(r)};{\bm\theta}))}{\partial {\bm h}_{t}^{(r)}}
    \frac{\partial {\bm h}_{t}^{(r)}}{\partial {\bm h}_{k}^{(r)}}
    \frac{\partial {\bm h}_{k}^{(r)}}{\partial {\bm\theta}}
\end{equation}
The partial derivatives on the right hand side can be efficiently computed using the back-propagation through time (BPTT) algorithm \cite{werbos1990backpropagation}. The partial derivative in the middle, $\partial{\bm h}_{t}^{(r)}/\partial {\bm h}_{k}^{(r)}$, represents the dependencies between time steps $t$ and $k$. Unfortunately, this partial derivative is known to exhibit stability issues as the difference between time steps increases \cite{bengio1994learning}.
Since BPTT development in the 1980s \cite{werbos1990, rumelhart1985, robinson1987}, it has inspired several advancements aimed at improving temporal modelling using neural networks and speeding up gradient computation \cite{pascanu2013, Hochreiter1997, cho2014learning, vaswani2017, williams1990, williams1989, graves2014}. However, limited work has been done on improving parallelization. 

\section{Fixed-Point Representation}
\label{section:Fixed-Point Representation}
Consider applying the Elman style RNN in~\cref{eq:elman} to a sequence of length 3. Starting from the initial history vector ${\bm h}_{0}$ set to some value, the history states at each subsequent time can be computed as follows
\begin{equation}
\begin{matrix}
    {\bm h}_{0} & = & {\bm h}_{0}\hspace{1.9cm}\\
    {\bm h}_{1} & = & {\bm\phi}({\bm W}{\bm h}_{0} + {\bm V}{\bm x}_{0})\\
    {\bm h}_{2} & = & {\bm\phi}({\bm W}{\bm h}_{1} + {\bm V}{\bm x}_{1})\\
    {\bm h}_{3} & = & {\bm\phi}({\bm W}{\bm h}_{2} + {\bm V}{\bm x}_{2})
\end{matrix}
    \label{eq.RNN}
\end{equation}
Noting that the history states ${\bm h}_{0}$, ${\bm h}_{1}$, ${\bm h}_{2}$, ${\bm h}_{3}$ appear on the left hand side {\em and} on the right hand side (ignoring for the moment the lack of ${\bm h}_{3}$ on the right hand side and the misalignment of indices) suggests that these equations can be written as a fixed-point equation $\vec{\bm h}={\bm f}(\vec{\bm h},\vec{\bm x})$. Indeed, it is possible to show that the above set of equations can be written in the following fixed-point form
\begin{equation}
    \vec{\bm h} = \vec{\bm\phi}(\vec{\bm W}\vec{\bm h} + \vec{\bm V} \vec{\bm x})
    \label{eq.elman vector}
\end{equation}
where
\begin{eqnarray}
    \vec{\bm x}&=&[{\bm 0}^\top\;{\bm x}_{0}^\top\;\ldots\;{\bm x}_{T-1}^\top]^{\top}\\
    \vec{\bm h}&=&[{\bm h}_{0}^{\top}\;\ldots\;{\bm h}_{T}^{\top}]^\top\\ \vec{\bm\phi}(\vec{\bm z})&=&[{\bm z}_{0}^{\top}\;{\bm\phi}({\bm z}_{1})^{\top}\;\ldots\;{\bm\phi}({\bm z}_{T})^{\top}]^\top
\end{eqnarray}
and
\begin{equation}
    \vec{\bm W} = 
    \begin{bmatrix}
        {\bm I}\! & \!{\bm 0}\! & \!\ldots\! & \!{\bm 0}\\
        {\bm W}\! & \!{\bm 0}\! & \!\ldots\! & \!{\bm 0}\\
        \vdots\!  & \!\ddots\! & \!\vdots\! & \!\vdots\\
        {\bm 0}\! & \!\ldots\! & \!{\bm W}\! & \!{\bm 0}
    \end{bmatrix}, 
    \vec{\bm V} = 
    \begin{bmatrix}
        {\bm 0}\! & \!{\bm 0}\! & \!\ldots\! & \!{\bm 0}\\
        {\bm 0}\! & \!{\bm V}\! & \!\ldots\! & \!{\bm 0}\\
        \vdots\!  & \!\vdots\!  & \!\ddots\! & \!\vdots\\
        {\bm 0}\! & \!{\bm 0}\! & \!\ldots\! & \!{\bm V}
    \end{bmatrix}
\end{equation}
For a sequence with three input words ($T=3$) the fixed-point form in~\cref{eq.elman vector} can be expressed as
\begin{equation}
\!\!\!\begin{bmatrix}
    {\bm h}_0\\
    {\bm h}_1\\
    {\bm h}_2\\
    {\bm h}_3
\end{bmatrix}
\!\!=\!
\vec{\bm\phi}\left(\!
\begin{bmatrix}
    {\bm I} \!&\! {\bm 0} \!&\! {\bm 0} \!&\! {\bm 0}\\
    {\bm W} \!&\! {\bm 0} \!&\! {\bm 0} \!&\! {\bm 0}\\
    {\bm 0} \!&\! {\bm W} \!&\! {\bm 0} \!&\! {\bm 0}\\
    {\bm 0} \!&\! {\bm 0} \!&\! {\bm W} \!&\! {\bm 0}
\end{bmatrix}
\!\!\!\!
\begin{bmatrix}
    {\bm h}_0\\
    {\bm h}_1\\
    {\bm h}_2\\
    {\bm h}_3
\end{bmatrix}
\!\!+\!\!
\begin{bmatrix}
    {\bm 0} \!&\! {\bm 0} \!&\! {\bm 0} \!&\! {\bm 0}\\
    {\bm 0} \!&\! {\bm V} \!&\! {\bm 0} \!&\! {\bm 0}\\
    {\bm 0} \!&\! {\bm 0} \!&\! {\bm V} \!&\! {\bm 0}\\
    {\bm 0} \!&\! {\bm 0} \!&\! {\bm 0} \!&\! {\bm V}
\end{bmatrix}
\!\!\!\!
\begin{bmatrix}
    {\bm 0}\\
    {\bm x}_0\\
    {\bm x}_1\\
    {\bm x}_2
\end{bmatrix}
\!\right)
\end{equation}
It is easy to verify that this equation simplifies to~\cref{eq.RNN}.
Thus, the sequence of RNN history states for any sequence is the fixed-point of the non-linear equation system in~\cref{eq.elman vector}. Non-linear equation systems appear in a range of machine learning approaches \cite{zhou2018graph, chen2018neural, bai2019}. 
However, it is rarely possible to obtain exact fixed-points of those equations and closed-forms in particular. Instead, approximate fixed-points are commonly used. 

\subsection{Fixed-point iteration (FPI) algorithm}
The Banach theorem \cite{zhou2018graph} 
states that regardless of the initial starting point $\vec{\bm h}^{(0)}$ the following iterative process $\vec{\bm h}^{(n+1)} = {\bm f}(\vec{\bm h}^{(n)},\vec{\bm x})$ will converge exponentially fast to the fixed-point of $\vec{\bm h}={\bm f}(\vec{\bm h},\vec{\bm x})$. For the fixed-point form in~\cref{eq.elman vector} the corresponding iterative process can be written as 
\begin{equation}
\vec{\bm h}^{(n+1)} = \vec{\bm\phi}(\vec{\bm W}\vec{\bm h}^{(n)} + \vec{\bm V} \vec{\bm x})
\label{eq.elman vector update}
\end{equation}
One interesting aspect of this process is that it enables {\em parallel} computation of all history states. 
Another interesting aspect is that it will converge to $\vec{\bm h}=[{\bm h}_{0}^{\top}\;\ldots\;{\bm h}_{T}^\top]^\top$ (RNN history states) exactly in $T$ steps. To illustrate this, consider the same example with the sequences of length 3 above. Starting with $\vec{\bm h}^{(0)} = [{\bm h}_{0}^{(0)}\; {\bm h}_{1}^{(0)}\; {\bm h}_{2}^{(0)}\; {\bm h}_{3}^{(0)}]^\top$, where ${\bm h}_{0}^{(0)}$ is set to the initial RNN state ${\bm h}_{0}$ and others set to arbitrary values, the first update based on equation~\eqref{eq.elman vector update} yields
\begin{equation}
    \label{eq.3-words-h1}
    \vec{\bm h}^{(1)} = 
    \begin{bmatrix}
    {\bm h}_{0}^{(0)}\\
        {\bm\phi}({\bm W}{\bm h}_{0}^{(0)}+{\bm V}{\bm x}_{0})\\
        {\bm\phi}({\bm W}{\bm h}_{1}^{(0)}+{\bm V}{\bm x}_{1})\\
        {\bm\phi}({\bm W}{\bm h}_{2}^{(0)}+{\bm V}{\bm x}_{2})
    \end{bmatrix}
\end{equation}
The second update will yield
\begin{equation}
    \label{eq.3-words-h2}
    \vec{\bm h}^{(2)} = 
    \begin{bmatrix}
    {\bm h}_{0}^{(0)}\\
    \bm\phi({\bm W}{\bm h}_{0}^{(0)}+{\bm V}{\bm x}_{0})\\
    \bm\phi({\bm W}\bm\phi({\bm W}{\bm h}_{0}^{(0)}+{\bm V}{\bm x}_{0})+{\bm V}{\bm x}_{1})\\
    \bm\phi({\bm W}\bm\phi({\bm W}{\bm h}_{1}^{(0)}+{\bm V}{\bm x}_{1})+{\bm V}{\bm x}_{2})
    \end{bmatrix}
\end{equation}
Finally, the third update will yield
\begin{equation}
    \label{eq.3-words-h3}
    \vec{\bm h}^{(3)} \!\!=\!\!
    \begin{bmatrix}
    {\bm h}_{0}^{(0)}\\
        {\bm\phi}({\bm W}\underbrace{{\bm h}_{0}^{(0)}}_{{\bm h}_{0}}+{\bm V}{\bm x}_{0})\\
        \bm\phi({\bm W}\underbrace{\bm\phi({\bm W}{\bm h}_{0}^{(0)}+{\bm V}{\bm x}_{0})}_{{\bm h}_1}+{\bm V}{\bm x}_{1})\\
        \bm\phi({\bm W}\underbrace{\bm\phi({\bm W}\bm\phi({\bm W}{\bm h}_{0}^{(0)}\!\!+\!{\bm V}{\bm x}_{0})\!+\!{\bm V}{\bm x}_{1})}_{{\bm h}_2}\!+\!{\bm V}{\bm x}_{2})\!
    \end{bmatrix}
    \!\!\!\equiv\!\!\!
    \begin{bmatrix}
    {\bm h}_{0}\\
    {\bm h}_{1}\\
    {\bm h}_{2}\\
    {\bm h}_{3}
    \end{bmatrix}
\end{equation}
which is equivalent to the RNN history states. Note that all initial history states other than ${\bm h}_{0}^{(0)}$ have been eliminated. Although the parallel forward pass with the fixed-point representation yields identical to the RNN history states values, the parallel backward pass would lead to some dependencies counted multiple times. For example, as shown in~\cref{eq.3-words-h3} the first word ${\bm x}_{0}$ will make 3 contributions, the second word ${\bm x}_{1}$ will make 2 contributions and the final word ${\bm x}_{2}$ will make 1 contribution to the gradient. Thus, the information coming further from the past would receive more boosting than the more recent information. 

\subsection{Approximate fixed points (AFP)}
\label{ssec:fpr_afp}
The Banach theorem \cite{zhou2018graph} states that the intermediate points $\vec{\bm h}^{(n)}$ converge to the fixed-point $\vec{\bm h}$ exponentially fast, which lends support for approximating fixed-points by terminating the iterative process after a small number of iterations $\rho$. The section below illustrates a pseudo-code for updating the RNN parameters in the $e$-th training epoch (mini-batch size is 1 for simplicity), where $\vec{\bm y}=[{\bm y}_{0}^{\top}\;{\bm y}_{1}^{\top}\ldots{\bm y}_{T}^{\top}]^\top$ is a sequence of output probability distributions, $\vec{\bm U}$ is a block-diagonal matrix with diagonal elements set to ${\bm U}$ and $\vec{\bm\sigma}$ is a block-softmax function. 
\begin{algorithm}[h]
\caption{Approximate fixed-point training}
\label{algorithm:MRNN}
\For{$r$ in $[1,R]$}{
    \For{$n$ in $[1,\rho]$}{
        $\vec{\bm h}^{(r,n)} = \vec{\bm\phi}(\vec{\bm W}\vec{\bm h}^{(r,n-1)} + \vec{\bm V}\vec{\bm x}^{(r)})$\\
        {\color{blue}\small// optionally skip propagation of dependencies}\\
        {\color{blue}$\vec{\bm h}^{(r,n)} \!\!=\!\! \left.\vec{\bm h}^{(r,n)}\right|_{{\bm\theta}={\bm\theta}^{(e)}}$\!\!\!\!\!\!{}}\!\!\!\!\!\!\!\!\!\!\!
    }
    $\vec{\bm y}^{(r)} = \vec{\bm\sigma}(\vec{\bm U}{\bm h}^{(r,\rho)})$\\
    Accumulate loss\\
}
Update parameters
\end{algorithm}
The simple example in \cref{eq.3-words-h1,eq.3-words-h2,eq.3-words-h3} shows that each iteration adds dependencies on one more past word. Thus, the iteration number $\rho$ is akin to the order of Markov assumption. The limited nature of dependencies possible with the approximate fixed-points make them suitable for lattice rescoring. By expanding lattices to order $\rho+1$ enables fixed-points to be applied to all order $\rho+1$ arcs in parallel. In contrast to RNNs, which require inference time approximations, the approximate fixed-points are consistent in both training and inference.


Although approximate fixed points (AFP) can be computed in parallel, as the example in  \cref{eq.3-words-h1,eq.3-words-h2,eq.3-words-h3} shows the overall amount of computation performed will be larger than in the standard RNN case. Table~\ref{Table: complexity} compares complexities in RNN and FPI based forward passes, where $T$ is the length of input sequence, $H$ is the size of the history state, 
$C$ is the number of computations (addition, multiplications, etc.) in the history state and $\rho$ is the number of iterations. 
\begin{table}[!htbp]
    \centering
    \caption{Comparison of complexities in recurrent neural network and approximate fixed point based forward passes}
    \vspace{-0.2cm}
    \begin{tabular}{c|c|c|c}
    \toprule
    \!\!\!\!Optimisation &
    \!\!\!\begin{tabular}[c]{@{}c@{}}Time\\complexity\end{tabular}\!\!\! &
    \begin{tabular}[c]{@{}c@{}}Space\\ complexity\end{tabular} &
    \begin{tabular}[c]{@{}c@{}}Computational\\ complexity\end{tabular} \\
    \midrule
    \!\!\!\!BPTT    & \!\!\!\!${\mathcal O}(T)$ \!\!\!\!&  \!\!${\mathcal O}(HT)$\!\!   \!\!& ${\mathcal O}(CT)$\!\!  \\
    \!\!\!\!FPI     & \!\!\!\!${\color{red}{\mathcal O}(\rho)}$ \!\!\!\!&  \!\!${\mathcal O}(2HT)$ \!\!  & \!\!${\mathcal O}(CT\rho)$\!\! \\
    \bottomrule
    \end{tabular}
    \label{Table: complexity}
\end{table}
As mentioned above the overall amount of computation in the backward pass is also larger in the RNN case. To reduce the amount of computation in the backward pass it is possible to eliminate the propagation of dependencies from the past completely as shown in the line ${\bm 5}$ of the Algorithm~\ref{algorithm:MRNN} above. Note that the lack of dependency propagation does not affect the computed history state values but will cause differences in the computed gradients.%
%
%
\section{Experiments}
\label{section:Experiments}
The experiments conducted in this work focused on assessing approximate fixed-points (AFP) in two language modelling tasks. The following Section~\ref{ssec:exp_setup} provides details about the configuration of baselines and AFPs. Sections~\ref{ssec:exp_ptb} and \ref{ssec:exp_wiki} then report on their performance in each task respectively.
%
\subsection{Experimental setup}
\label{ssec:exp_setup}
The PennTree Bank (PTB) \cite{marcus1993} and the WikiText-2 \cite{merity2016pointer} used in this work are two relatively small scale datasets. Table~\ref{tab:dataset comparison} provides basic statistics including the percentage of out-of-vocabulary (OOV) words mapped to the special `word` {\tt <unk>}. 
%
\vspace{-0.2cm}
\begin{table}[htbp]
\caption{Basic statistics of PTB and wikitext2 datasets}
\vspace{-0.2cm}
\label{tab:dataset comparison}
\centering
\begin{tabular}{cccc}
\toprule
\multirow{2}*{Dataset} & \multicolumn{2}{c}{Training words} & OOV (\%)   \\
 & Unique & Total & (train/dev/test)\\
\midrule
PTB         & 10K & 888K & 5.1 / 5.0  / 6.1 \\ 
WikiText-2  & 33K & 2M   & 2.7 / 5.5 / 6.3\\ 
\bottomrule
\end{tabular}
\end{table}\\
\textbf{PTB:} This dataset has been often used for conducting experiments in language modelling \cite{mikolov2010}. 
The original dataset consists of sentences extracted from the Wall Street Journal corpus \cite{paul1992design}. 
A pre-processed version, which provides 888 thousand training words and 10 thousand word vocabulary, is used in this work. The development and test subsets contain 70 and 79 thousand words respectively. Compared with modern datasets, the vocabulary size in this dataset is small.\\
\textbf{WikiText-2:} This dataset consists of articles extracted from Wikipedia. A pre-processed version available as a part of PyTorch toolkit \cite{paszke2017automatic} 
is used in this work. Compared to the PTB, the training set and vocabulary sizes in this dataset are approximately 2 and 3 times larger respectively. Note that this dataset contains many foreign (e.g. Japanese) words which contribute to the increased vocabulary size. 
%

A number of simple (feed-forward NN (FFNN)) and more complex (RNN, Long Short-Term Memory (LSTM))
baselines have been investigated. 
%
%
All NN LMs were implemented in PyTorch. The minibatch size used in stochastic optimisation was 20. The optimisation was performed using Adam \cite{kingma2014adam} 
for either 20 (PTB) or 40 (WikiText-2) epochs. The FFNN baseline is a bigram LM, which lacks information about all previous words other than the most recent one when making predictions. The initial history states of all recurrent LMs (including AFP) were set to zeros. Unless stated otherwise, the size of history states is 100. All other parameters were initialised randomly using the uniform distribution ${\mathcal U}(-0.05, 0.05)$. AFPs were configured identical to the RNN baselines. The estimation of AFPs benefited from the sparsity of $\vec{\bm W}$, $\vec{\bm V}$ and $\vec{\bm U}$ matrices.
%
\subsection{Penn Tree Bank}
\label{ssec:exp_ptb}
\begin{figure}[htbp]
\centering
\begin{tikzpicture}
\begin{axis}[xlabel=Iteration ($\rho$), ylabel=PPL, height=5cm, width=8.5cm, x label style={at={(axis description cs:0.5,0.02)},anchor=north}, y label style={at={(axis description cs:0.05,.5)}},]
\addplot[mark=*, black] coordinates {(1,142.10) (15,142.10)};
\addlegendentry{BPTT}
\addplot[mark=square*,red] table [x=iteration,y=fpi] {plot/ptb.tex};
\addlegendentry{FPI}
\addplot[mark=square,blue] table [x=iteration,y=cfpi] {plot/ptb.tex};
\addlegendentry{FPI$|_{{\bm\theta}={\bm\theta}^{(e)}}$}
\end{axis}
\end{tikzpicture}
\vspace{-0.6cm}
\caption{PPL performance in the PTB task against the AFP iteration number $\rho$}
\label{fig:fpi_ptb}
\end{figure}
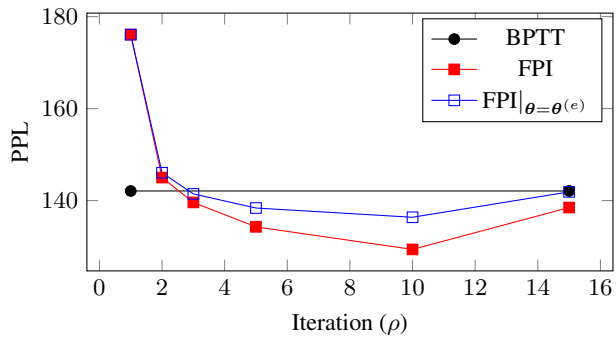

The first experiment examined learning approximate fixed points (AFP) using the fixed point iteration (FPI) algorithm in the PTB task. As discussed in Section~\ref{ssec:fpr_afp}, the FPI algorithm can be configured to either propagate dependencies or not (FPI$|_{{\bm\theta}={\bm\theta}^{(e)}}$) during the training. Figure~\ref{fig:fpi_ptb} compares perplexities (PPL) of these two configurations against the number of iterations $\rho$ used by the FPI algorithm.
Note that in this task the average number of words per sentence is 20. As Figure~\ref{fig:fpi_ptb} shows, just two iterations of both FPI algorithms bring AFP performance close to that obtained with the BPTT-trained RNN that has access to the complete past information. Additional FPI iterations yield more gains for AFP. As expected, the removal of dependencies in FPI$|_{{\bm\theta}={\bm\theta}^{(e)}}$ has a negative effect on the final performance. 
Table~\ref{tab:PTB} puts these results in the context of baselines other than the BPTT-trained RNN.  
\begin{table}[htbp]
\centering
\caption{Summary of perplexities in the PTB task}
\vspace{-0.2cm}
\label{tab:PTB}
\begin{tabular}{clc}
\toprule
LM & Optimisation & PPL \\
\midrule
5-gram \cite{mikolov2012context} & Kneser-Ney & 141.2\\
FFNN$_2$ & BP & 179.0\\
RNN \cite{mikolov2012context} & TBPTT & 142.1\\
RNN & BPTT & 142.1\\
\hline
\multirow{2}*{AFP} & FPI & 129.4\\
 & FPI$|_{{\bm\theta}={\bm\theta}^{(e)}}$ & 136.4\\
\bottomrule
\end{tabular}
\end{table}
The FFNN$_2$ (bigram) baseline provides an interesting contrast. Similar to FPI$|_{{\bm\theta}={\bm\theta}^{(e)}}$ it does not extract statistics from words more than 1 time step in the past. However, the former benefits from the information accumulated within history states.%
\subsection{WikiText-2}
\label{ssec:exp_wiki}
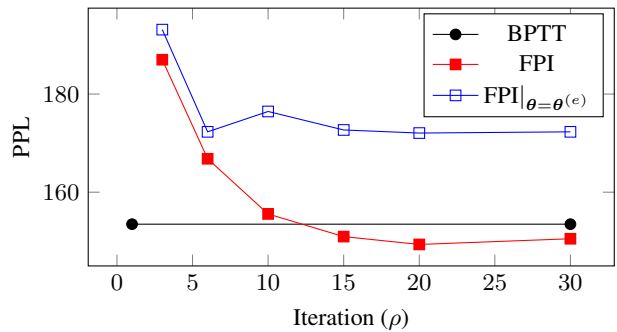
\begin{figure}[htbp]
\centering
\begin{tikzpicture}
\begin{axis}[xlabel=Iteration ($\rho$), ylabel=PPL, height=5cm, width=8.5cm, x label style={at={(axis description cs:0.5,0.02)},anchor=north}, y label style={at={(axis description cs:0.05,.5)}},]
\addplot[mark=*, black] coordinates {(1,153.48) (30,153.48)};
\addlegendentry{BPTT}
\addplot[mark=square*,red] table [x=iteration,y=fpi] {plot/wiki.tex};
\addlegendentry{FPI}
\addplot[mark=square,blue] table [x=iteration,y=cfpi] {plot/wiki.tex};
\addlegendentry{FPI$|_{{\bm\theta}={\bm\theta}^{(e)}}$}
\end{axis}
\end{tikzpicture}
\vspace{-0.6cm}
\caption{PPL performance in the WikiText-2 task against the AFP iteration number $\rho$}
\label{fig:fpi_wiki}
\end{figure}

Many high-performance recipes for WikiText-2 task involve the use of large, multi-layer, recurrent models. Due to a limited computational resource available, a simpler PTB-like configuration was created. Table~\ref{tab:wiki} shows the impact of simplifications applied to the initial LSTM model with 2 hidden layers and 1024 units per layer trained using the truncated BPTT \cite{williams1990efficient}
and dropout \cite{srivastava2014dropout}.
\begin{table}[htbp]
\centering
\caption{Summary of perplexities in the WikiText-2 task}
\vspace{-0.2cm}
\label{tab:wiki}
\begin{tabular}{ccclc}
\toprule
LM & Layers & Units & Optimisation & PPL \\
\midrule
\multirow{4}*{LSTM} & \multirow{3}*{2} & 1024 & \multirow{2}*{TBPTT+dropout} & 99.3\\
\cline{3-3}
 & 2 & \multirow{3}*{200} & & 108.59\\
\cline{4-5}
 & & & \multirow{2}*{TBPTT} & 131.18\\
\cline{2-2}
 & 1 & & & 132.32\\
\midrule
\multirow{2}*{RNN} & \multirow{4}*{1} & \multirow{4}*{200} & TBPTT & 168.31\\
 & & & BPTT & 153.48\\
\cline{1-1}\cline{4-5}
\multirow{2}*{AFP} & & & FPI & 149.35\\ 
 & & & FPI$|_{{\bm\theta}={\bm\theta}^{(e)}}$ & 172.05\\ 
\bottomrule
\end{tabular}
\end{table}
Compared to the more advanced baseline, the final RNN model with single hidden layer and 200 units as expected shows a significantly worse performance yet is computationally more affordable. Figure~\ref{fig:fpi_wiki} repeats the investigation conducted in the PTB task (see Figure~\ref{fig:fpi_wiki}). 
Note that in this task the average number of words per sentence is 55, which is more than 2.5 times longer than in the PTB task. The AFP requires more iterations to match the performance of BPTT-trained RNNs, which suggests that this task requires significantly longer context for making accurate predictions. The latter would also provide an explanation why the simpler FPI$|_{{\bm\theta}={\bm\theta}^{(e)}}$ variant performs worse in this task. Overall these results illustrate that AFPs can offer competitive performance to BPTT-trained RNNs. 
%
\section{Conclusion and Future Work}
\label{section:Conclusion and Future Work}
This paper proposed a novel view on history representations obtained from recurrent neural networks (RNN) as fixed-points of non-linear equation systems. The novel view gives rise to a fixed-point iteration (FPI) algorithm that enables approximate but accurate history representations to be obtained in constant rather than linear with respect to sequence length time. The approximate fixed-points (AFP) enable efficient training (parallelization) and provide an opportunity for consistent inference on complex structures, such as lattices, which is impractical with RNNs trained using standard algorithms such as back-propagation through time (BPTT). Experimental validation was performed in two language modelling tasks where AFPs achieved competitive performance to BPTT-trained RNNs given a small number of FPI iterations. The future work will examine applying AFPs to lattice rescoring and will explore more advanced AFP optimisation.
\bibliographystyle{IEEEtran}
\bibliography{reference}

\begin{thebibliography}{10}
\providecommand{\url}[1]{#1}
\csname url@samestyle\endcsname
\providecommand{\newblock}{\relax}
\providecommand{\bibinfo}[2]{#2}
\providecommand{\BIBentrySTDinterwordspacing}{\spaceskip=0pt\relax}
\providecommand{\BIBentryALTinterwordstretchfactor}{4}
\providecommand{\BIBentryALTinterwordspacing}{\spaceskip=\fontdimen2\font plus
\BIBentryALTinterwordstretchfactor\fontdimen3\font minus
  \fontdimen4\font\relax}
\providecommand{\BIBforeignlanguage}[2]{{%
\expandafter\ifx\csname l@#1\endcsname\relax
\typeout{** WARNING: IEEEtran.bst: No hyphenation pattern has been}%
\typeout{** loaded for the language `#1'. Using the pattern for}%
\typeout{** the default language instead.}%
\else
\language=\csname l@#1\endcsname
\fi
#2}}
\providecommand{\BIBdecl}{\relax}
\BIBdecl

\bibitem{sutskever2014}
I.~Sutskever, O.~Vinyals, and Q.~V. Le, ``Sequence to sequence learning with
  neural networks,'' in \emph{NIPS}, 2014.

\bibitem{mikolov2010}
T.~Mikolov, M.~Karafi\'{a}t, L.~Burget, J.~\v{C}ernock\'{y}, and S.~Khudanpur,
  ``Recurrent neural network based language model,'' in \emph{Interspeech},
  2010.

\bibitem{graves2012}
A.~Graves, ``Sequence transduction with recurrent neural networks,''
  \emph{arXiv:1211.3711}, 2012.

\bibitem{chan2016}
W.~Chan, N.~Jaitly, Q.~Le, and O.~Vinyals, ``Listen, attend and spell: A neural
  network for large vocabulary conversational speech recognition,'' in
  \emph{ICASSP}, 2016.

\bibitem{peddinti2017low}
V.~Peddinti, Y.~Wang, D.~Povey, and S.~Khudanpur, ``Low latency acoustic
  modeling using temporal convolution and {LSTM}s,'' \emph{IEEE Signal
  Processing Letters}, vol.~25, no.~3, pp. 373--377, 2017.

\bibitem{pascanu2013}
R.~Pascanu, T.~Mikolov, and Y.~Bengio, ``On the difficulty of training
  recurrent neural networks,'' in \emph{ICML}, 2013.

\bibitem{Hochreiter1997}
S.~Hochreiter and J.~Schmidhuber, ``Long short-term memory,'' \emph{Neural
  computation}, 1997.

\bibitem{cho2014learning}
K.~Cho, B.~Van~Merri{\"e}nboer, C.~Gulcehre, D.~Bahdanau, F.~Bougares,
  H.~Schwenk, and Y.~Bengio, ``Learning phrase representations using {RNN}
  encoder-decoder for statistical machine translation,'' \emph{arXiv preprint
  arXiv:1406.1078}, 2014.

\bibitem{vaswani2017}
A.~Vaswani, N.~Shazeer, N.~Parmar, J.~Uszkoreit, L.~Jones, A.~N. Gomez,
  L.~Kaiser, and I.~Polosukhin, ``Attention is all you need,'' in \emph{NIPS},
  2017.

\bibitem{schwenk2012continuous}
H.~Schwenk, ``Continuous space translation models for phrase-based statistical
  machine translation,'' in \emph{COLING}, 2012, pp. 1071--1080.

\bibitem{bengio1994learning}
Y.~Bengio, P.~Simard, and P.~Frasconi, ``Learning long-term dependencies with
  gradient descent is difficult,'' \emph{IEEE Transactions on Neural Networks},
  vol.~5, no.~2, pp. 157--166, 1994.

\bibitem{chen2014efficient}
X.~Chen, Y.~Wang, X.~Liu, M.~J. Gales, and P.~C. Woodland, ``Efficient
  {GPU}-based training of recurrent neural network language models using
  spliced sentence bunch,'' in \emph{Interspeech}, 2014.

\bibitem{werbos1990backpropagation}
P.~J. Werbos, ``Backpropagation through time: what it does and how to do it,''
  \emph{Proceedings of the IEEE}, vol.~78, no.~10, pp. 1550--1560, 1990.

\bibitem{williams1990efficient}
R.~J. Williams and J.~Peng, ``An efficient gradient-based algorithm for on-line
  training of recurrent network trajectories,'' \emph{Neural computation},
  vol.~2, no.~4, pp. 490--501, 1990.

\bibitem{liu2014efficient}
X.~Liu, Y.~Wang, X.~Chen, M.~J.~F. Gales, and P.~C. Woodland, ``Efficient
  lattice rescoring using recurrent neural network language models,'' in
  \emph{ICASSP}, 2014, pp. 4908--4912.

\bibitem{xu2018pruned}
H.~Xu, T.~Chen, D.~Gao, Y.~Wang, K.~Li, N.~Goel, Y.~Carmiel, D.~Povey, and
  S.~Khudanpur, ``A pruned {RNNLM} lattice-rescoring algorithm for automatic
  speech recognition,'' in \emph{ICASSP}, 2018.

\bibitem{zhou2018graph}
J.~Zhou, G.~Cui, Z.~Zhang, C.~Yang, Z.~Liu, L.~Wang, C.~Li, and M.~Sun, ``Graph
  neural networks: A review of methods and applications,'' \emph{arXiv preprint
  arXiv:1812.08434}, 2018.

\bibitem{elman1990finding}
J.~L. Elman, ``Finding structure in time,'' \emph{Cognitive science}, 1990.

\bibitem{jordan1997serial}
M.~I. Jordan, ``Serial order: A parallel distributed processing approach,'' in
  \emph{Advances in psychology}.\hskip 1em plus 0.5em minus 0.4em\relax
  Elsevier, 1997, vol. 121, pp. 471--495.

\bibitem{BiRNN1997}
M.~{Schuster} and K.~K. {Paliwal}, ``Bidirectional recurrent neural networks,''
  \emph{IEEE Transactions on Signal Processing}, 1997.

\bibitem{pennington2014glove}
J.~Pennington, R.~Socher, and C.~D. Manning, ``Glove: Global vectors for word
  representation,'' in \emph{EMNLP}, 2014, pp. 1532--1543.

\bibitem{mikolov2013efficient}
T.~Mikolov, K.~Chen, G.~Corrado, and J.~Dean, ``Efficient estimation of word
  representations in vector space,'' \emph{arXiv preprint arXiv:1301.3781},
  2013.

\bibitem{bottou2010large}
L.~Bottou, ``Large-scale machine learning with stochastic gradient descent,''
  in \emph{COMPSTAT}.\hskip 1em plus 0.5em minus 0.4em\relax Springer, 2010,
  pp. 177--186.

\bibitem{werbos1990}
P.~J. Werbos, ``Backpropagation through time: what it does and how to do it,''
  \emph{{IEEE}}, 1990.

\bibitem{rumelhart1985}
D.~E. Rumelhart, G.~E.~H. GE, and R.~J. Williams, ``Learning internal
  representations by error propagation,'' California University San Diego,
  Tech. Rep., 1985.

\bibitem{robinson1987}
A.~J. Robinson and F.~Fallside, ``The utility driven dynamic error propagation
  network,'' Cambridge University, Tech. Rep., 1987.

\bibitem{williams1990}
R.~J. Williams and J.~Peng, ``An efficient gradient-based algorithm for on-line
  training of recurrent network trajectories,'' \emph{Neural Computation},
  1990.

\bibitem{williams1989}
R.~J. Williams and D.~Zipser, ``A learning algorithm for continually running
  fully recurrent neural networks,'' \emph{Neural Computation}, 1989.

\bibitem{graves2014}
A.~Graves, G.~Wayne, and I.~Danihelka, ``Neural {Turing} machines,''
  \emph{arXiv:1410.5401}, 2014.

\bibitem{chen2018neural}
R.~T. Chen, Y.~Rubanova, J.~Bettencourt, and D.~Duvenaud, ``Neural ordinary
  differential equations,'' \emph{arXiv preprint arXiv:1806.07366}, 2018.

\bibitem{bai2019}
S.~Bai, J.~Z. Kolter, and V.~Koltun, ``Deep equilibrium models,'' in
  \emph{NIPS}, 2019.

\bibitem{marcus1993}
M.~P. Marcus, B.~Santorini, and M.~A. Marcinkiewicz, ``Building a large
  annotated corpus of {E}nglish: The {P}enn {T}reebank,'' \emph{Computational
  Linguistics}, 1993.

\bibitem{merity2016pointer}
S.~Merity, C.~Xiong, J.~Bradbury, and R.~Socher, ``Pointer sentinel mixture
  models,'' \emph{arXiv preprint arXiv:1609.07843}, 2016.

\bibitem{paul1992design}
D.~B. Paul and J.~Baker, ``The design for the {W}all {S}treet {J}ournal-based
  {CSR} corpus,'' in \emph{Speech and Natural Language}, Harriman, New York,
  1992.

\bibitem{paszke2017automatic}
A.~Paszke, S.~Gross, S.~Chintala, G.~Chanan, E.~Yang, Z.~DeVito, Z.~Lin,
  A.~Desmaison, L.~Antiga, and A.~Lerer, ``{Automatic Differentiation in
  PyTorch},'' in \emph{NIPS Workshop on Autodiff}, 2017.

\bibitem{kingma2014adam}
D.~P. Kingma and J.~Ba, ``Adam: A method for stochastic optimization,''
  \emph{arXiv preprint arXiv:1412.6980}, 2014.

\bibitem{mikolov2012context}
T.~Mikolov and G.~Zweig, ``Context dependent recurrent neural network language
  model,'' in \emph{2012 IEEE Spoken Language Technology Workshop (SLT)}.\hskip
  1em plus 0.5em minus 0.4em\relax IEEE, 2012.

\bibitem{srivastava2014dropout}
N.~Srivastava, G.~Hinton, A.~Krizhevsky, I.~Sutskever, and R.~Salakhutdinov,
  ``Dropout: a simple way to prevent neural networks from overfitting,''
  \emph{The Journal of machine learning research}, 2014.

\end{thebibliography}
\end{document}